\documentstyle[12pt]{article}
\begin{document}
\begin{center}
{\bf CANONICAL TRANSFORMATIONS AND SOLDERING}\\
 \vskip 2cm
R. Banerjee \footnote {e-mail address: rabin@boson.bose.res.in } \\
\vskip 1cm
S. N. Bose National Centre for Basic Sciences,\\
Block JD, Sector III, Salt Lake, Calcutta 700091, India.\\
\vskip 1cm
and\\
\vskip 1cm
 Subir Ghosh \footnote {e-mail address: subir@boson.bose.res.in ; sghosh@isical.ac.in} \\ \vskip 1cm
P. A. M. U., Indian Statistical Institute,\\
203 B. T. Road, Calcutta 700035, India.\\
\end{center}
\vskip 3cm
Abstract:\\
We show that the recently developed soldering formalism in the Lagrangian approach and canonical transformations in the Hamiltonian approach are complementary. The examples of gauged chiral bosons in two dimensions and self-dual models in three dimensions are discussed in details.

\newpage
The concept of soldering has proved extremely useful in different contexts. The soldering formalism essentially combines two distinct Lagrangians manifesting dual aspects of some symmetry (like left-right chirality or self and anti-self duality etc.) to yield a new Lagrangian which is devoid of, or rather hides, that symmetry. The quantum interference effects, whether constructive \cite{abw} or destructive \cite{w}, among the dual aspects of symmetry, are thereby captured through this mechanism. Alternatively, in the Hamiltonian formulation, Canonical Transformations (CT) can be sometimes used to decompose a composite Hamiltonian into two distinct pieces. A familiar example \cite{djt} is the decomposition of the Hamiltonian of a particle in two dimensions, moving in a constant magnetic field and quadratic potential, into two pieces corresponding to the Hamiltonians of two one dimensional oscillators, rotating in a clockwise and an anti-clockwise direction, respectively. It a!
ppears, therefore, that the solder

ing formalism which fuses the symmetries while CT which decouples the symmetries are complementary to each other. In the present paper, we shall elaborate on these notions by considering a particular symmetry, namely {\it chirality}. This is known to play a pivotal role in discussing different aspects of two dimensional field theories with left movers (leftons) and right movers (rightons), ideas which  have also been used in string theory context \cite{gw}.

Consider, eg. two such Lagrangians 
$L_+(q_+,~\dot q_+)$ and $L_-(q_-,~\dot q_-)$, that permit a soldering. The basic variables transform identically under a transformation
\begin{equation}
\delta q_+=\delta q_-=\alpha .
\label{eqst}
\end{equation}
The main idea is to show that although $L_\pm $ are not invariant under these transformations, it is possible to devise a modified Lagrangian
\begin{equation}
L(q_\pm, \dot q_\pm, \eta )= L_+(q_+,\dot q_+)+L_-(q_-,\dot q_-)+\Delta (q_\pm,\dot q_\pm, \eta),
\label{eqls}
\end{equation}
that will be invariant.
The new (external) field $\eta $ is the soldering field which can be eliminated in favour of the original variables by using the equations of motion. An explicit form for $\Delta $ is obtained such that $L$ remains invariant under the soldering transformation. The soldered Lagrangian incidentally, no longer depends on $q_\pm $, but on their difference $q_+-q_-=q $. Hence, this Lagrangian is manifestly invariant under the transformations (\ref{eqst}).

The Hamiltonian obtained from the soldered Lagrangian by a formal Legendre transformation is denoted in terms of its canonical pairs by $H(q,p)$. Performing a CT into a new  canonical set $(Q,P)$, the Hamiltonian is changed to $H(Q,P)$. For systems containing the dual aspects of some symmetry, this $H(Q,P)$ actually decomposes into distinct pieces,
\begin{equation}
H(Q,P)=H_1(q_1,p_1)+H_2(q_2,p_2),
\label{eqh}
\end{equation}
where the new set $(Q,P)$ consistis of independent canonical pairs $(q_1,p_1)$ and $(q_2,p_2)$. Indeed, as we will show later, the matching of the degrees of freedom count between the original and final system is crucial.
It is now possible to identify these pieces with $H_\pm $ obtained from the original $L_\pm $, thereby establishing a connection between the soldering formulation and CT. 

Before going to field theory in two dimensions, let us first consider quantum mechanics in these dimensions where the basic ideas are illuminated in a simple way. A very familiar example, alluded earlier, is provided by the quantum mechanical model \cite{djt},
\begin{equation}
L={m\over 2}\dot x_i^2+{B\over 2}\epsilon_{ji}x_j\dot x_i-{K\over 2}x_i^2;~~~i=1,2
\label{eqlq}
\end{equation}
which describes the planar motion of a unit charged particle in a constant magnetic field $B$ and a prescribed electric field. The Hamiltonian is given by
\begin{equation}
H=p^i\dot x_i-L={1\over{2m}}(p_i+{B\over 2}\epsilon_{ij}x_j)^2+{K\over 2}x_i^2,
\label{eqhq}
\end{equation}
with $p^i={{\partial L}\over{\partial \dot x_i}}$. Performing a CT \cite{djt},
$$p_+={\sqrt{{w_+}\over{2m\Omega}}}p_1+{\sqrt{{w_+m\Omega}\over 2}}x_2;~~~
p_-={\sqrt{{w_-}\over{2m\Omega}}}p_1-{\sqrt{{w_-m\Omega}\over 2}}x_2,$$
\begin{equation}
x_+={\sqrt{{m\Omega }\over{2\omega _+}}}x_1-{\sqrt{1\over{2w_+m\Omega}}}p_2;~~~
x_-={\sqrt{{m\Omega }\over{2\omega _-}}}x_1+{\sqrt{1\over{2w_-m\Omega}}}p_2,
\label{eqct}
\end{equation}
where
\begin{equation}
w_\pm=\Omega\pm {B\over {2m}},~~~\Omega={\sqrt{{{B^2}\over{4m^2}}+{K\over m}}}
\label{ct1}\end{equation}
 the Hamiltonian takes the form
\begin{equation}
H=H_++H_-={1\over 2}[p_+^2+w_+^2x_+^2]+
{1\over 2}[p_-^2+w_-^2x_-^2].
\label{eqhh1}
\end{equation}
This corresponds to the Hamiltonian of two decoupled Harmonic Oscillators with independent canonical pairs $(x_+,p_+)$ and $(x_-,p_-)$, respectively.

The above analysis can be understood strictly in the Lagrangian formalism by following our soldering prescription \cite{bk}. The Hamiltonians 
$H_\pm $, in (\ref{eqhh1}) can be derived from the following Lagrangians respectively,
\begin{equation}
L_+={1\over 2}(w_+\epsilon_{ij}x_i\dot x_j-w_+^2x_i^2)~;~~~
L_-={1\over 2}(-w_-\epsilon_{ij}y_i\dot y_j-w_-^2y_i^2)~,
\label{eqll1}
\end{equation}
which have a non-trivial algebra, following from their symplectic structure,
$$\{x_i,~x_j\}=-{1\over {w_+}}\epsilon_{ij};~~~\{y_i,~y_j\}={1\over {w_-}}\epsilon_{ij}.$$
These characterise one dimensional oscillators rotating in the 
clockwise and anti-clockwise sense with frequencies $w_+$ and $w_-$ respectively. Hence $L_+$ and $L_-$ can be soldered as shown in \cite{bk,bg}. In fact, $L_\pm $ mimic the left and right movers 
,(or leftons and rightons), which one usually associates with chiral field theory models in two dimensional space time. The basic steps of soldering are just recapitulated. Consider the transformations,
\begin{equation}
\delta x_i=\delta y_i=\eta_i.
\label{eqxy}
\end{equation}
It is possible to construct a modified Lagrangian, \cite{bk}
\begin{equation}
L=L_+(x_i)+L_-(y_i)+W_i[J_i^+(x_i)+J_i^-(y_i)]-{1\over 2}(w_+^2+w_-^2)W_i^2,
\label{eqls1}
\end{equation}
with
$$J_{\pm i}(z_i)=w_\pm (\pm\dot z_i+w_\pm \epsilon_{ij}z_j);~~~z_i=x_i,~y_i,$$
which is invariant under the transformations (\ref{eqxy}) together with $\delta W_i=\epsilon_{ij}\eta _j$. Eliminating the soldering field $W_i$ from  (\ref{eqls1}) we obtain the final Lagrangian in terms of the difference of original variables,
\begin{equation}
L={1\over 2}\dot X_i^2+{1\over 2}(w_--w_+)\epsilon_{ij}X_i\dot X_j-{1\over 2}w_+w_-X_i^2 ;~~X_i={\sqrt {{w_+w_-}\over{w_+^2+w_-^2}}}
(x_i-y_i).
\label{eqx}
\end{equation}
With the identifications,
$$~~~w_--w_+=-{B\over m},~~~w_-w_+={K\over m}$$
which follow from (\ref{ct1}),
the above Lagrangian goes over to (\ref{eqlq}). This shows the dual roles of soldering and CT complementing each other. It is however essential that the oscillators must have the left-right symmetry (as in (\ref{eqll1}) to effect the soldering. Observe that if $w_+=w_-$, then (\ref{eqx}) just reduces to the Lagrangian of a two-dimensional oscillator. Physically speaking, two one dimensional chiral oscillators moving in opposite directions have been combined to yield a conventional planar oscillator. If, however, $w_+\ne w_-$, the left and right oscillators do not cancel so that a net rotational motion survives. This is the origin of the generation of the "magnetic field" effect in (\ref{eqx}).

Let us now consider two dimensional field theory. 
An explicit one loop computation 
 of the two dimensional chiral fermion determinant in the presence of an external abelian gauge field, yields \cite{aar}, in the bosonised language, the following results,
\begin{equation}
W_\pm={1\over{4\pi}}\int d^2x(\partial_+\phi \partial_-\phi +2eA_\pm\partial_\mp\phi +ae^2A_+A_-).
\label{eqwpm}
\end{equation}
where we have introduced light cone variables,
$$A_\pm={1\over{\sqrt 2}}(A_0\pm A_1)=A^\mp;~~\partial_\pm={1\over{\sqrt 2}}(\partial_0\pm \partial_1)=\partial^\mp,$$
and $a$ is a parameter manifesting bosonization or regularization ambiguities. Note that our regularization preserves Bose symmetry \cite{bb}, so that the same factor $a$ appears in either expression. The soldering of $W_+(\phi )$ with $W_-(\rho )$  is easily done \cite{abw} by exploiting the relevant chiral symmetries. Consider the transformation,
\begin{equation}
\delta \phi=\delta \rho=\alpha;~~\delta A_\pm=0.
\label{eqst1}
\end{equation}
Introducing the soldering fields $B_\pm $, it is possible to verify that the modified effective action,
\begin{equation}
W[\phi, \rho ]=W_+[\phi ]+W_-[\rho ]-\int d^2x[B_-J_+(\phi )+B_+J_-(\rho)]
+{1\over{2\pi}}\int d^2x B_+B_-,
\label{eqws}
\end{equation}
with the currents,
$$J_\pm(\eta)={1\over{2\pi }}(\partial_\pm \eta+eA_\pm);~~\eta=\phi , \rho ,$$
is invariant under the symmetry including (\ref{eqst1}) and $\delta B_\pm=\partial_\pm\alpha $. Eliminating $B_\pm $, by using the equations of motion, the soldered effective action is given by
\begin{equation}
W[\theta ]={1\over{4\pi }}\int d^2x(\partial_+\theta \partial_-\theta +2eA_+\partial _-\theta-2eA_-\partial_+\theta +2(a-1)e^2A_+A_-),
\label{eqwb}
\end{equation}
where $\theta =\phi -\rho $. Conventional gauge invariance is restored for $a=1$. Thus with this particular value, we see how the individual components pertaining to the left and right chiral effective actions are soldered to yield the gauge invariant result for the vector effective action. This is an example of a constructive interference. If we had included the conventional Maxwell (gauge field) term, then this analysis shows how the two massless modes of the chiral models $(a=1)$  are fused to yield the single massive mode of the vector Schwinger model \cite{abw}.

The above analysis has an exact analogue in the Hamiltonian formulation based on CTs. The gauge invariant $(a=1)$ Lagrangian for the vector theory (\ref{eqwb}) is reexpressed as,
\begin{equation}
W={1\over{4\pi }}\int d^2x{\cal L};~~~
{\cal L}={1\over 2}(\dot\theta ^2-\theta '^2)+{\sqrt 2} e[A_+(\dot\theta -\theta ')-A_-(\dot\theta +\theta ')].
\label{eqll}
\end{equation}
The corresponding Hamiltonian is
\begin{equation}
{\cal H}={1\over 2}[(\pi -eA_++eA_-)^2+\theta '^2]+e(A_++A_-)\theta ',
\label{eqhh}
\end{equation}
where we have scaled $\sqrt 2 e\equiv e$ and $\pi ={{\partial {\cal L}}\over {\partial \dot \theta }}$. The Hamiltonian in (\ref{eqhh}), under the following CT,
$$\theta '={1\over 2}(-\theta _1 '+\pi _ 1+\theta_ 2 '+\pi_ 2),$$
\begin{equation}
\pi={1\over 2}(\theta_ 1 '-\pi_ 1 +\theta_ 2' +\pi _2 ),
\label{eqctt}
\end{equation}
where $(\theta _1, \pi _1)$ and $(\theta _2, \pi _2 )$ are the new canonical pairs, gets decoupled and is given by
$${\cal H}=[({1\over 2}\theta _1 '-{1\over 2}\pi _1 )^2-2eA_+({1\over 2}\theta _1 '-{1\over 2}\pi _1)+{{e^2}\over 2}(A_+^2-A_+A_-)]$$
\begin{equation}
+[({1\over 2}\theta _2 '+{1\over 2}\pi _2 )^2+2eA_-({1\over 2}\theta _2 '+{1\over 2}\pi _2)+{{e^2}\over 2}(A_-^2-A_+A_-)]={\cal H} _R+{\cal H}_L,
\label{eqhrl}
\end{equation}
where the independent pieces are
\begin{equation}
{\cal H}_R=[(\psi _R' )^2-2eA_+\psi _R'+{{e^2}\over 2}(A_+^2-A_+A_-)],
\label{eqr3}
\end{equation}
\begin{equation}
{\cal H}_L=
[(\psi _L')^2+2eA_-\psi _L'+{{e^2}\over 2}(A_-^2-A_+A_-)].
\label{eqhr2}
\end{equation}
Here we have identified 
$$\psi _L'\equiv {1\over 2}\theta _2 '+{1\over 2}\pi _2,~~~ \psi _R' \equiv {1\over 2}\theta _1 '-{1\over 2}\pi _1, $$  
which corresponds to the basic algebra,
\begin{equation}
\{\psi _L(x),\psi _L(y)\}={1\over 2}\delta '(x-y),~~~
\{\psi _R(x),\psi _R(y)\}=-{1\over 2}\delta '(x-y).
\label{eqca}
\end{equation}
This is nothing but the well known left and right chiral boson algebra \cite{sfr}.
It is seen that the original Hamiltonian has decoupled into two distinct pieces which are identified with the left and right gauged chiral boson Hamiltonians \cite{h}, associated with a gauged lefton and righton, respectively.

A word about the degree of freedom count may be useful. The CT has decomposed the original boson to two (left and right) chiral bosons. A single degree of freedom in configuration space is thus shown to consist of two times half a degree of freedom, also in configuration space.

On the other hand, we now apply the above CTs to the original Lagrangians  ${\cal L}_+(\phi )$ and ${\cal L}_-(\rho )$, defined from (\ref{eqwpm}), that were being soldered,
$${\cal L}_-(\rho )={1\over 2}(\dot \rho ^2-\rho '^2)+eA_-(\dot \rho + \rho ')+{{e^2}\over 2}A_+A_-,$$
\begin{equation}
{\cal L}_+(\phi )={1\over 2}(\dot \phi ^2-\phi '^2)+eA_+(\dot \phi - \phi ')+{{e^2}\over 2}A_+A_-.
\label{eqlch}
\end{equation}
Again we have scaled $\sqrt 2 e\equiv e$.
The corresponding Hamiltonians are
$${\cal H}_-(\rho )={1\over 2}(\pi ^\rho -eA_-)^2+{1\over 2}\rho '^2-eA_-\rho '-{{e^2}\over 2}A_+A_-,$$
\begin{equation}
{\cal H}_+(\phi )={1\over 2}(\pi ^\phi -eA_+)^2+{1\over 2}\phi '^2+eA_+\phi '-{{e^2}\over 2}A_+A_-.
\label{eqhpm}
\end{equation}
 Applying CTs similar to (\ref{eqctt}),
$$\phi '={1\over 2}(-\phi _1 '+\pi _1^\phi +\phi _2 '+\pi_2^\phi );~~
\pi ^\phi ={1\over 2}(\phi _1 '-\pi _1 ^\phi +\phi _2 '+\pi_2 ^\phi),$$
$$\rho '={1\over 2}(-\rho _1 '+\pi _1 ^\rho +\rho _2 '+\pi_2 ^\rho);~~
\pi ^\rho ={1\over 2}(\rho _1 '-\pi _1 ^\rho +\rho _2 '+\pi_2 ^\rho ),$$
on each of the Hamiltonians ${\cal H}_\pm$ we find,
\begin{equation}
{\cal H}_+
=[(\eta _R')^2-2eA_+\eta _R'+{{e^2}\over 2}(A_+^2-A_+A_-)]+(\eta _L')^2,
\label{eqh1}
\end{equation}
\begin{equation}
{\cal H}_-
=[(\chi _L')^2-2eA_-\chi _L'+{{e^2}\over 2}(A_-^2-A_+A_-)]+(\chi _R')^2.
\label{eqh2}
\end{equation}
The fields are identified as,
\begin{equation}
\eta _R'\equiv {1\over 2}\phi_1'-{1\over 2}\pi_1 ^\phi,~~\eta _L'\equiv
{1\over 2}\phi_2'+{1\over 2}\pi_2^\phi ;~~~\chi _R'\equiv {1\over 2}\rho_1'-{1\over 2}\pi_1 ^\rho ,~~\chi _L'\equiv {1\over 2}\rho _2'+{1\over 2}\pi_2 ^\rho .
\label{eqid}
\end{equation}
Note that $\eta _L $, $\chi _L $ satisfy the lefton algebra while $\eta _R $, $\chi _R $ satisfy the righton algebra given in (\ref{eqca}).
Both ${\cal H}_+$ and ${\cal H}_-$
 have split up into two components such that there is a free chiral boson and an interacting one of opposite chirality. Ignoring the free chiral component the interacting ones exactly match with the gauged chiral components 
in (\ref{eqr3},\ref{eqhr2}), with the following identification  $\eta _R\equiv \psi _R,~~-\chi _L\equiv \psi _L $. This is the central result of our paper showing how the soldering mechanism in the Lagrangian formalalism and the CT in the Hamiltonian formalism are connected. The degree of freedom count exactly parallels the analysis given earlier.  

A more direct contact between the soldering formulation and CT is also possible
in this context. The Lagrangians (\ref{eqlch}) that were originally soldered have, in their interaction, 
only one chiral component. Thus, as far as the interactions are concerned, the effective degree of 
freedom is only half, the other half being free. This is more clearly seen in the structure
of the hamiltonians in (\ref{eqh1}, \ref{eqh2}). Thus when the Lagrangians (\ref{eqlch}) were soldered to yield (\ref{eqll}), the single degree 
of freedom associated with the interaction was revealed, while the free part did not manifest itself.
This is nothing wrong since the contribution from the free Lagrangian can always be
absorbed in the normalisation of the path integral. However, it is possible to directly start from 
gauged chiral boson Lagrangians, whose degree of freedom is exactly half. The extra 
half degree of freedom assocaited with the free part is non-existant. These Lagrangians are \cite{h},
\begin{equation}
{\cal L}_R=-\dot \phi \phi '-\phi '^2 -2e\phi '(A_0+A_1)-{{e^2}\over 2}(A_0+A_1)^2 +{{e^2}\over 2}A_\mu A^\mu,
\label{eqch1}
\end{equation}
\begin{equation}
{\cal L}_L=\dot \rho \rho '-\rho '^2 +2e\rho '(A_0-A_1)-{{e^2}\over 2}(A_0-A_1)^2 +{{e^2}\over 2}A_\mu A^\mu,
\label{eqch2}
\end{equation}
which precisely correspond to the Hamiltonians ${\cal H}_R$ and ${\cal H}_L$ given in (\ref{eqr3},\ref{eqhr2}). We now show that the soldering of ${\cal L}_R$  with ${\cal L}_L$ yields (\ref{eqll}). Taking the variations (\ref{eqst1}) we find,
\begin{equation}
\delta {\cal L}_R=2J_R\alpha ',~~~\delta {\cal L}_L=2J_L\alpha ',
\label{eqxyz}
\end{equation}
where the currents are
\begin{equation}
J_R=-(\dot\phi +\phi '+e(A_0+A_1)),~~J_L=(\dot\psi -\psi '+e(A_0-A_1)).
\label{eqxx}
\end{equation}
Introducing the soldering field $B$, which transformed as,
\begin{equation}
\delta B=-2\alpha '
\label{eqzz}
\end{equation}
it is easy to check that the Lagrangian
\begin{equation}
{\cal L}={\cal L}_R+{\cal L}_L+B(J_R+J_L)-{1\over 2}B^2,
\label{eqp}
\end{equation}
is invariant under the combined transformations (\ref{eqst1}) and (\ref{eqzz}). Eliminating $B$ in favour of the other variables yields the soldered Lagrangian. This is exactly (\ref{eqll}) with the basic field defined as $\theta =\phi -\rho $.

As a final illustration, which would be a field theoretic extension of the  model in (\ref{eqlq}), consider the self-dual models in 2+1-dimensions \cite{dj},
\begin{equation}
{\cal L}_\pm(h)={1\over 2}h_\mu h^\mu\pm{1\over{2m_\pm}}\epsilon_{\mu\nu\lambda}h^\mu \partial ^\nu h^\lambda;~~~h=f,~~g.
\label{eqsd}
\end{equation}
A straightforward analysis yields the following  field algebras and Hamiltonians (with appropriate renaming of variables),
$$\{f_1(x),~~f_2(y)\}=-m_+\delta(x-y),$$
$${\cal H}_+={1\over 2}f_i^2+{1\over{2m_+^2}}(\epsilon_{ij}\partial _if_j)^2={1\over 2}(\pi _f^2+m_+^2f^2)+{1\over{2m_+^2}}(\epsilon_{ij}\partial _if_j)^2~;$$
and,
$$\{g_1(x),~~g_2(y)\}=m_-\delta(x-y),$$
\begin{equation}
{\cal H}_-={1\over 2}g_i^2+{1\over{2m_-^2}}(\epsilon_{ij}\partial _ig_j)^2=
{1\over 2}(\pi _g^2+m_-^2g^2)+{1\over{2m_-^2}}(\epsilon_{ij}\partial _ig_j)^2.
\label{db}
\end{equation}
Following the usual steps of soldering of the Lagrangians ${\cal L}_+(f)$ and ${\cal L}_-(g)$
in (\ref{eqsd}) we get the soldered Lagrangian as \cite{bk},
\begin{equation}
{\cal L}={1\over 2}A_\mu A^\mu -{\theta \over{2m^2}}\epsilon_{\mu\nu\lambda}\partial^\mu A^\nu A^\lambda -{1\over{4m^2}}A_{\mu\nu}A^{\mu\nu},
\label{eqcs}
\end{equation}
with $$A_{\mu\nu }=\partial_\mu A_\nu -\partial_\nu A_\mu;~~~
A_\mu\equiv f_\mu -g_\mu ;~~~m_+-m_-=\theta ;~~~m_+m_-=m^2.$$
Going over to the Hamiltonian formulation, the first step is to obtain the canonical Hamiltonian,
\begin{equation}
{\cal H}=\Pi ^i\dot A_i-{\cal L}=
{{m^2}\over 2}\Pi_i \Pi_i +({1\over 2}+{{\theta ^2}\over{8m^2}})A_iA_i-{\theta\over 2}\epsilon_{ij}\Pi_iA_j +{1\over {4m^2}}A_{ij}A_{ij}.
\label{eqhcs}
\end{equation}
where $\Pi ^i={{\partial {\cal L}}\over {\partial \dot A_i}}$. Due to the presence of spatial derivatives, it is problematic to decouple the $A_{ij}A_{ij}$ term. This may be contrasted with the Maxwell theory where such a decoupling in terms of Harmonic Oscillators in the momentum space is possible only after a proper choice of gauge, (in particular the Coulomb gauge) \cite{iz}. Since the present theory is not a gauge theory, the above mechanism fails. We thus work in the approximation where the term $A_{ij}A_{ij}$ can be neglected. In other words, we are looking at the long wave length limit and keep the smallest number of derivatives. Going over to a new set of independent canonical variables,
\begin{equation}
\{A_-(x),~~\Pi_-(y)\}=\delta(x-y);~~~\{A_+(x),~~\Pi_+(y)\}=-\delta(x-y),
\label{eqaa}
\end{equation}
by the following CT,
$$A_\pm={1\over 4}{\sqrt {{m_++m_-}\over{m_\mp}}}(\Pi_1\mp {{m_++m_-}\over {2m_+m_-}}A_2);$$
\begin{equation}
\Pi_\pm=4{\sqrt {{m_\mp}\over{m_++m_-}}}
(-{{m_+m_-}\over {m_++m_-}}\Pi_2\mp {1\over 2}A_1),
\label{eqct1}
\end{equation}
the Hamiltonian decouples into
\begin{equation}
{\cal H}={1\over 2}[(\Pi_-^2+m_-^2A_-^2)+(\Pi_+^2+m_+^2A_+^2)].
\label{eqhh2}
\end{equation}
Each of the pieces is now mapped to the previously obtained Hamiltonians of the self and anti-self dual models in the long wavelength limit, using the following identifications,
$$g_2\equiv \Pi_-,~g_1\equiv A_-;~~~f_1\equiv \Pi_+,~f_2\equiv -A_+~.$$

The soldering formalism, as elaborated here, is applicable only for Lagrangians manifesting dual aspects of some symmetry. Exploiting this feature, it is possible to combine these Lagrangians to yield a new Lagrangian. Canonical Transformations, on the other hand, can be performed on any Hamiltonian. However, the effect of the Canonical Transformation to decouple the Hamiltonian into distinct and independent pieces is essentially tied to the dual aspects of the symmetry. The roles of the two mechanisms is therefore complementary, which has been amply illustrated here. Apart from this, the canonical transformations given here provide an alternative way of gauging chiral bosons without the necessity of any ad-hoc insertions of constraints \cite{h}. Since the study of gauged chiral bosons has been revived \cite{aw}, such an approach might be useful.

\end{document}